\documentclass[aps,prd,preprint,showpacs,nofootinbib]{revtex4}
\usepackage{epsfig}

\begin{document}

\title{Next-to-Leading Order QCD Corrections to the Direct Top Quark Production
via Model-independent FCNC Couplings at Hadron Colliders}

\author{Jian Jun Liu\footnote{\hspace{-0.1cm}Present address:
Physics department, Tsinghua University, Beijing 100084, China;
Electronics address: liujj@mail.tsinghua.edu.cn}, Chong Sheng
Li\footnote{\hspace{-0.1cm}Electronics address: csli@pku.edu.cn},
and Li Lin Yang} \affiliation{Department of Physics, Peking
University, Beijing 100871, China}

\author{Li Gang Jin}
\affiliation{Institute of Theoretical Physics, Academia Sinica,
Beijing 100080, China}

\date{\today}

\begin{abstract}
We calculated the next-to-leading order (NLO) QCD corrections to
the cross sections for direct top quark productions induced by
model--independent flavor changing neutral current couplings at
hadron colliders. The NLO results increase the experimental
sensitivity to the anomalous couplings. Our results show that the
NLO QCD corrections enhance the leading order (LO) total cross
sections at the Tevatron Run 2 about $60\%$ for both of
$\kappa_{tc}^g$ and $\kappa_{tu}^g$ couplings, and enhance the LO
total cross sections at the LHC about $40\%$ for $\kappa_{tc}^g$
couplings and $50\%$ for $\kappa_{tu}^g$ couplings, respectively.
Moreover, the NLO QCD corrections vastly reduce the dependence of
the total cross sections on the renormalization or factorization
scale, which leads to increased confidence in predictions based on
these results.
\end{abstract}


\maketitle

\section{Introduction}
The flavor dynamics, such as the mixing of three generation
fermions and their large mass differences observed, is still a
great mystery in particle physics today. The heaviest one of three
generation fermions is the top quark with the mass close to the
electroweak (EW) symmetry breaking scale. Therefore, it can play
the role of a wonderful probe for the EW breaking mechanism and
new physics beyond the standard model (SM) through its decays and
productions. An important aspect of the top quark physics is to
investigate anomalous flavor changing neutral current (FCNC)
couplings. Within the SM, the FCNC processes are forbidden at the
tree-level and highly suppressed by the Glashow-Iliopoulos-Maiani
mechanism \cite{GIM} at one-loop level. All of the current precise
measurements of various FCNC processes, for example, $b
\rightarrow s\gamma$ \cite{pdg}, agree with the SM predictions
well. However, it should be noted that only few experimental
constraints on the top quark FCNC processes are available so far,
and many new physics models allow the existence of the tree-level
FCNC couplings, and may greatly enhance some FCNC processes. Since
the CERN Large Hadron Collider (LHC) can produce abundant top
quark events, the measurement of the top quark rare processes will
become possible. Therefore, a good understanding of the
theoretical predictions of top quark FCNC processes is important
for searching new physics.

Any new physics effect involved in top quark FCNC processes can be
incorporated into an effective Lagrangian in a model independent
way \cite{effect,beneke}:
\begin{eqnarray}
\mathcal{L}^{\mathrm{eff}} & = & -\frac{g}{2\cos\theta_W}
\sum_{q=u,c} \bar{t} \gamma^\mu (v_{tq}^Z -a_{tq}^Z\gamma_5)qZ_\mu
-\frac{g}{2\cos\theta_W} \sum_{q=u,c}
\frac{\kappa_{tq}^Z}{\Lambda} \bar{t}\sigma^{\mu\nu}(f_{tq}^Z
+ih_{tq}^Z\gamma_5)
qZ_{\mu\nu}  \nonumber \\
& &  -e \sum_{q=u,c}\frac{\kappa_{tq}^\gamma}{\Lambda}
\bar{t}\sigma^{\mu\nu} (f_{tq}^\gamma +ih_{tq}^\gamma\gamma_5)
qA_{\mu\nu} -g_s \sum_{q=u,c}\frac{\kappa_{tq}^g}{\Lambda}
\bar{t}\sigma^{\mu\nu} T^a (f_{tq}^g +ih_{tq}^g\gamma_5)
qG_{\mu\nu}^a  + \mathrm{h.c.}
\end{eqnarray}
where $\Lambda$ is the new physics scale, $\theta_W$ is the
Weinberg angle, and $T^a$ are the Gell-Mann matrices satisfying
${\rm Tr}\, (T^a T^b) = \delta^{ab}/2$. The $v^{Z}_{tq},
a^{Z}_{tq}$ can be complex in general, and $v^Z_{tq}=a^Z_{tq}=0$
in the SM. $G^a_{\mu \nu}=\partial _\mu G^a_\nu -
\partial _\nu G^a_\mu -g_s f^{abc} G_\mu^b G_\nu^c$ and
similarly for the other gauge bosons.
$\kappa$ is normalized to be real and positive and $f,h$ to be
complex numbers satisfying for each term $|f|^2 + |h|^2 = 1$.

The top quark FCNC processes induced by various anomalous
couplings have been studied in detail (for a most recent review,
see \cite{agui}). In general, top quark decay processes provide
the best place to discover top FCNC interactions involving
anomalous $t-q-\gamma$ and $t-q-Z$ couplings. However, for $t-q-g$
anomalous couplings, {\it i.e,} the forth term in Eq.(1), the best
processes are the top quark FCNC productions \cite{tait,young1},
and the direct top quark production processes are the most
sensitive ones \cite{young1}. The current constraints on the
$t-q-g$ anomalous couplings are $\frac{\kappa_{tq}^g}{\Lambda}\le
0.47\,{\rm TeV}^{-1}$ \cite{constraint}, and as pointed out in
\cite{young1,beneke} based on tree-level analysis of the direct
top quark productions, the couplings can further be detected down
to $\frac{\kappa_{tc}^g}{\Lambda}=0.0084\,{\rm TeV}^{-1}$,
$\frac{\kappa_{tu}^g}{\Lambda}=0.0033\,{\rm TeV}^{-1}$ at the LHC,
and $\frac{\kappa_{tc}^g}{\Lambda}=0.062\,{\rm TeV}^{-1}$,
$\frac{\kappa_{tu}^g}{\Lambda}=0.019\,{\rm TeV}^{-1}$ at the
Tevatron Run 2. However, since these processes involve strongly
interacting particles in their initial states and final states,
the lowest order results in general have large theoretical
uncertainties. In this paper, we present the complete NLO QCD
calculation for the cross section of direct top quark productions
induced by model--independent FCNC couplings at hadron colliders.

The arrangement of this paper is as follows. In Sec.~II we show
the LO results. In Sec.~III we present the details of the NLO
calculations including the virtual and real corrections. In
Sec.~IV by a numerical analysis we present the predictions for
total cross sections at hadron colliders.

\section{\label{sec:cal}The Leading Order Results}

We start from the flavor-changing operators including $t-q-g$
anomalous couplings:
\begin{equation}
-g_s \sum_{q=u,c}\frac{\kappa_{tq}^g}{\Lambda}
\bar{t}\sigma^{\mu\nu} T^a (f_{tq}^g +ih_{tq}^g\gamma_5)
qG_{\mu\nu}^a  + \mathrm{H.c.},
\end{equation}
and consider the leading order (LO) scattering process of hadrons
$P_1$ and $P_2$ to top quark:
\begin{eqnarray}
P_1(p_1)+P_2(p_2) \rightarrow g(k_1)+q(k_2)\rightarrow t(p_t)+X ~,
\ \ q=c\ {\rm or} \ u.
\end{eqnarray}

The spin and color averaged squared matrix element is
\begin{equation}
\overline{\left|{\mathcal{M}}^{\rm B}\right|^2} = \frac{8\pi
\alpha_s}{3} \left(\frac{\kappa_{tq}^g}{\Lambda}\right)^2 m_t^4,
\end{equation}
which, after the phase space integration, gives the LO partonic
cross section (the masses of charm quark and up quark are
neglected)
\begin{eqnarray}
\hat{\sigma}^{\rm B}(z) = \frac{8\pi^2 \alpha_s}{3s}
\left(\frac{\kappa_{tq}^g}{\Lambda}\right)^2m_t^2\delta (1-z),
\end{eqnarray}
where $s=(k_1+k_2)^2$ and $z=m_t^2/s$. The LO total cross section
at hadron colliders is obtained by convoluting the partonic cross
section with the parton distribution functions (PDFs) $G_{g,q/P}$
in the proton(anti-proton):
\begin{eqnarray}
\sigma^{\rm B}=\int dx_1dx_2
[G_{g/P_1}(x_1,\mu_F)G_{q/P_2}(x_2,\mu_F)+ (x_1\leftrightarrow
x_2)]\hat{\sigma}^{\rm B}.
\end{eqnarray}
where $\mu_F$ is the factorization scale.

\section{\label{sec:dis}The Next-to-Leading Order Calculations}
\begin{figure}[width=3.0textwidth]
\centerline{\epsfig{file=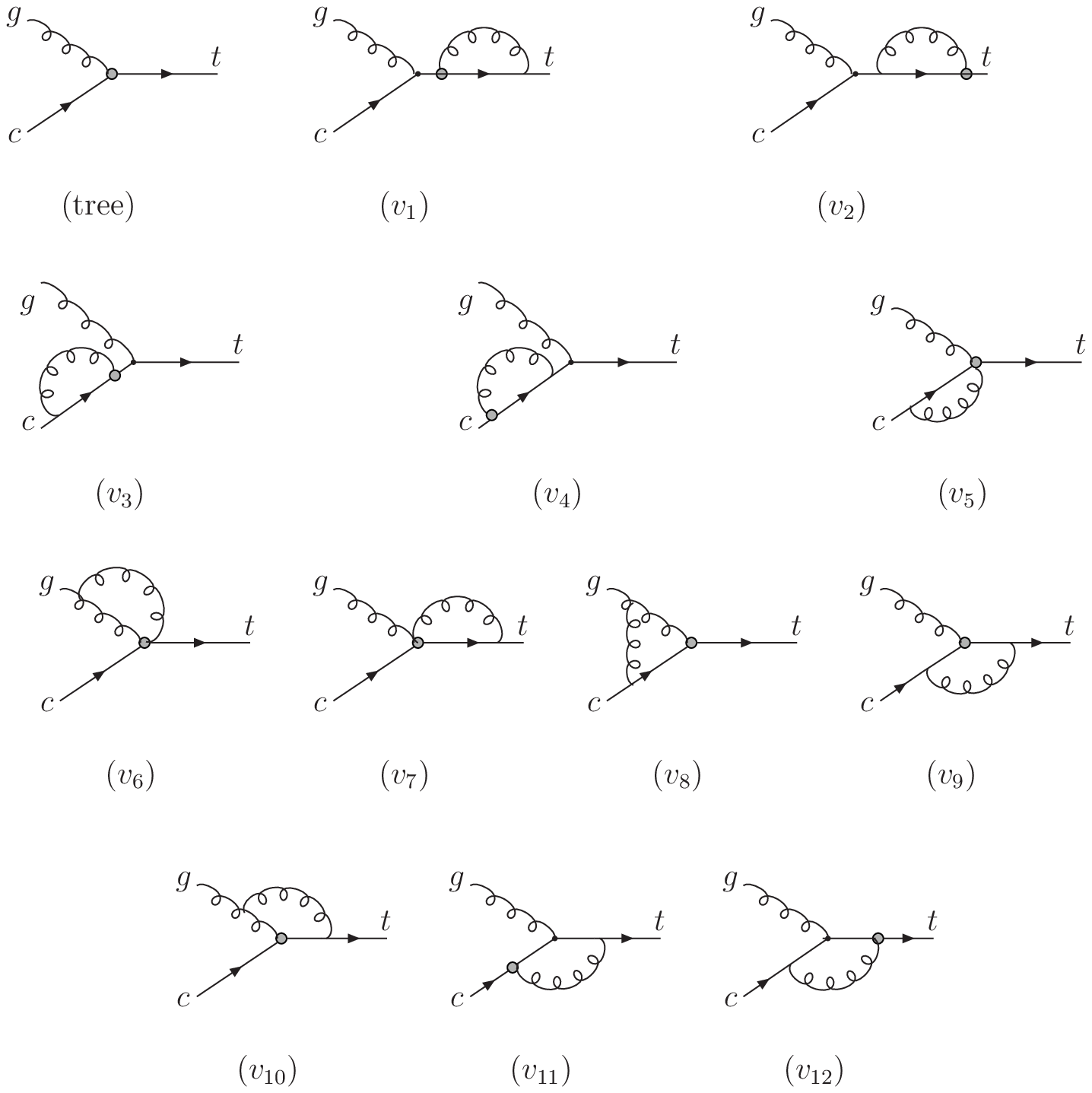,width=500pt}}
\caption[]{Tree-level and one-loop Feynman diagrams for the direct
top quark production. \label{fig:feynq}}
\end{figure}

\begin{figure}
\centerline{\epsfig{file=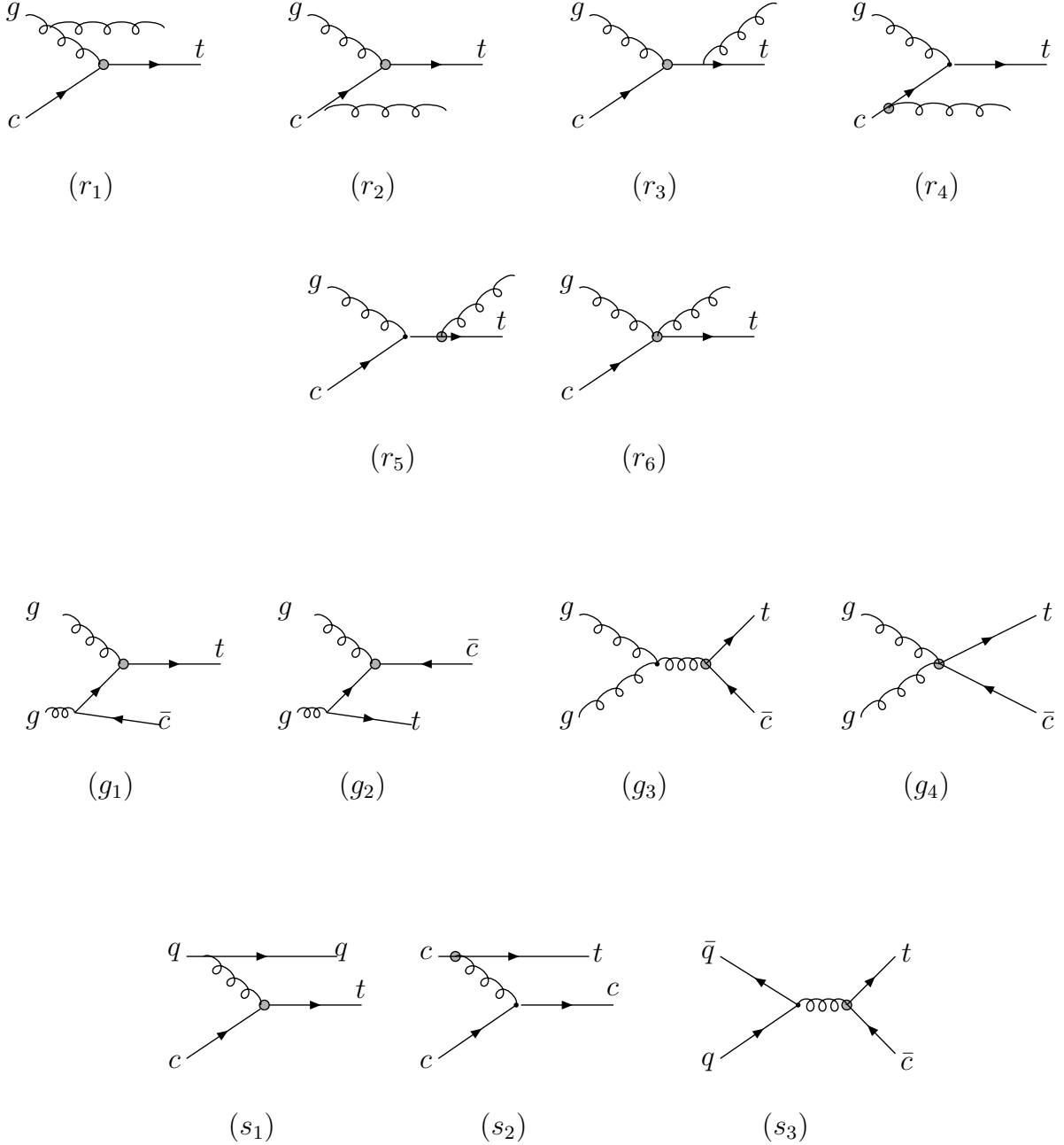,width=450pt}} \caption{Feynman
diagrams of real gluon emission sub-processes ($(r_1)-(r_6)$),
gluon initial state sub-processes ($(g_1)-(g_4)$), of which
($g_1$) and ($g_2$) have cross diagrams, and quark initial state
sub-processes ($(s_1)-(s_3)$), of which ($s_2$) has cross diagram.
\label{fig:feiynr}}
\end{figure}

For convenience, in this section we present only the results for
$gc\rightarrow t$, and the ones for $gu\rightarrow t$ can be
obtained by replacing $c$ with $u$ in related formulas. The cross
sections beyond LO involve the computation of one loop virtual
gluon corrections
[Figs.\ref{fig:feynq}($v_1$)--\ref{fig:feynq}($v_{12}$)] and real
gluon bremsstrahlung contributions to leading order processes
[Figs.\ref{fig:feiynr}($r_1$)--\ref{fig:feiynr}($r_6$)]. We also
include processes with two gluons in the initial states
[Figs.\ref{fig:feiynr}($g_1$)--\ref{fig:feiynr}($g_4$)] as well as
two quarks in the initial states
(Figs.\ref{fig:feiynr}($s_1$)--\ref{fig:feiynr}($s_3$)).  We will
encounter ultraviolet (UV) and infrared (IR) (soft and collinear)
divergences in our computation. We have used $n=4-2\epsilon$
dimensional regularization \cite{DREG} to regulate both these
divergences, and all divergences appear as $1/\epsilon^\alpha$
with $\alpha=1,2$.

The virtual corrections to direct top productions consist of
self-energy and vertex diagrams, which represent the QCD
corrections arising from quarks and gluons.  The total virtual
amplitude contains UV and IR divergences (the imaginary part is
neglected):
\begin{eqnarray}
{\mathcal{M}}^{{\rm virt}} &=& \frac{\alpha_s}{12\pi} C_\epsilon
\left(\frac{-13}{\epsilon_{IR}^2}
-\frac{17}{\epsilon_{IR}}+\frac{11}{\epsilon_{UV}}
+\frac{17\pi^2}{3}-15 \right) {\mathcal{M}}^{\rm B}
\nonumber \\[2ex] &&
+\left(\frac{1}{2}\delta Z_2^{(g)} +\frac{1}{2}\delta Z_2^{(c)}
+\frac{1}{2} \delta Z_2^{(t)} +\delta Z_{g_s} +\delta
Z_{\kappa^g_{tc}/\Lambda} \right) {\mathcal{M}}^{\rm B},
\end{eqnarray}
where $C_\epsilon=\Gamma(1+\epsilon)
\left(\frac{4\pi\mu_r^2}{m_t^2}\right)^{\epsilon}$, and the UV
divergences are renormalized by introducing counterterms for the
wave function of the external fields ($\delta Z_2^{(g)}$,$\delta
Z_2^{(c)}$, $\delta Z_2^{(t)}$), and the coupling constants
($\delta Z_{g_s}$,$\delta Z_{\kappa_{tc}^g/\Lambda}$). We define
these counterterms according to the following convention. For the
external fields, we fix $\delta Z_2^{(g)}$, $\delta Z_2^{(c)}$ and
$\delta Z_2^{(t)}$ using on-shell subtraction, and, therefore,
they also have IR singularities:
\begin{eqnarray}
\delta Z_2^{(g)} &=& -\frac{\alpha_s}{2\pi} C_\epsilon \left(
\frac{n_{f}}{3}-\frac{5}{2} \right) \left( \frac{1}{\epsilon_{UV}}
- \frac{1}{\epsilon_{IR}} \right) -\frac{\alpha_{s}}{6\pi}
C_\epsilon \frac{1}{\epsilon_{UV}}, \nonumber
\\
\delta Z_2^{(c)} &=& -\frac{\alpha_s}{3\pi} C_\epsilon \left(
\frac{1}{\epsilon_{UV}} - \frac{1}{\epsilon_{IR}} \right),
\nonumber
\\
\delta Z_2^{(t)} &=& -\frac{\alpha_s}{3\pi} C_\epsilon \left(
\frac{1}{\epsilon_{UV}} + \frac{2}{\epsilon_{IR}} +4 \right),
\end{eqnarray}
where $n_{f}=5$. For the renormalization of $g_s$, we use the
$\overline{\rm MS}$ scheme modified to decouple the top quark
\cite{colil}, i.e. the first $n_{f}$ light flavors are subtracted
using the $\overline{\rm MS}$ scheme, while the divergences
associated with the top-quark loop are subtracted at zero
momentum:
\begin{eqnarray}
\delta Z_{g_s} &=& \frac{\alpha_s}{4\pi}\Gamma(1+\epsilon)
(4\pi)^{\epsilon} (\frac{n_{f}}{3}-\frac{11}{2})
\frac{1}{\epsilon_{UV}} +\frac{\alpha_{s}}{12\pi}C_\epsilon
\frac{1}{\epsilon_{UV}}.
\end{eqnarray}
Thus, in this scheme, the renormalized strong coupling constant
$\alpha_s$ evolves with $n_{f}$ light flavors. Finally, for
$\delta Z_{\kappa_{tc}^g/\Lambda}$, we adopt the $\overline{\rm
MS}$ scheme and adjust it to cancel the remaining UV divergences
exactly:
\begin{eqnarray}
\delta Z_{\kappa_{tc}^g/\Lambda} &=& \frac{\alpha_s}{6\pi}
\Gamma(1+\epsilon) (4\pi)^{\epsilon} \frac{1}{\epsilon_{UV}}.
\end{eqnarray}
Therefore, now ${\mathcal{M}}^{{\rm virt}}$ has no UV-singular
contributions.

The partonic cross section of real gluon bremsstrahlung are
\begin{eqnarray}
\hat{\sigma}^{\rm real}(z,1/\epsilon_{IR})
&=&\frac{4\pi\alpha_s^2}{9s} \left(\frac{\kappa_{tc}^g}
{\Lambda}\right)^2m_t^2 C_\epsilon^\prime \Bigg[\left
(\frac{13}{\epsilon_{IR}^2} +\frac{4}{\epsilon_{IR}} \right)
\delta(1-z) +\frac{1}{\epsilon_{IR}} \Bigg(18z^2- 14z + 40 -{18
\over z} \nonumber
\\ [2ex]
&& -{26 \over (1-z)_+}\Bigg)+ 8\delta(1-z) +{1
\over 8(1-z)_+}\Bigg( 27z^2 -57z -2(77z + 27)\ln z -23  \nonumber
\\[2ex]
&& -{11 \over z} \Bigg) +\frac{77z+27}{2} \left({\ln(1-z) \over
1-z}\right)_+ -\left(36z^2 -28z+{83 \over 2}-{36 \over
z}\right)\ln(1-z) \nonumber
\\[2ex]
&& +\Bigg(28z^2 -22z +{107 \over 4}-{18 \over z}-{2 \over
1-z}\Bigg)\ln z +14 z^2-{73z \over 8}+{63 \over 4}-{109 \over 8z}
\Bigg],
\end{eqnarray}
where $C_\epsilon^\prime=\frac{\Gamma(1-\epsilon)}
{\Gamma(1-2\epsilon)} \left (\frac{4\pi\mu_r^2}{m_t^2}
\right)^{\epsilon}$. The ``plus" functions appearing in the above
results are the distributions which satisfy the following
equation:
\begin{eqnarray}
\int_0^1 dz~f_+(z)~ g(z) &=& \int_0^1 dz~f(z)
\Big(g(z)-g(1)\Big)\,, \label{eq43}
\end{eqnarray}
where $g(z)$ is any well-behaved function in the region $0\le z
\le 1$.

Combining the contributions of the LO result, the virtual
corrections and the real gluon bremsstrahlung, we obtain the bare
NLO partonic cross section:
\begin{eqnarray} \label{eq:gc}
\hat\sigma_{gc}^{\rm bare}(z,1/\epsilon_{IR}) &=& \frac{8\pi^2
\alpha_s}{3s} \left(\frac{\kappa_{tc}^g}
{\Lambda}\right)^2m_t^2\delta (1-z) \nonumber
\\[2ex] &&
+\frac{4\pi\alpha_s^2}{9s}
\left(\frac{\kappa_{tc}^g} {\Lambda}\right)^2m_t^2 \Bigg \{
C_\epsilon^\prime \frac{3}{\epsilon_{IR}} \Bigg[ -2N_c\left({z
\over (1-z)_+}+{1-z \over z} +z(1-z)\right) \nonumber
\\[2ex] &&
-\left({11 \over 2} - {n_f \over 3}\right) \delta(1-z) -
\frac{4}{3}\left({1+z^2 \over (1-z)_+}+{3 \over
2}\delta(1-z)\right) \Bigg] + \Bigg[\frac{4\pi^2}{3} -15 \nonumber
\\[2ex] &&
-\left (n_f -\frac{29}{2} \right) \ln\frac{\mu_r^2}{m_t^2}
\Bigg]\delta(1-z) +{1 \over 8(1-z)_+}\Bigg( 27z^2 -57z -2(77z +
27)\ln z \nonumber
\\[2ex] &&
-23 -{11 \over z} \Bigg) +\frac{77z+27}{2} \left({\ln(1-z) \over
1-z}\right)_+ -\left(36z^2 -28z+{83 \over 2}-{36 \over
z}\right)\ln(1-z) \nonumber
\\[2ex] &&
+\Bigg(28z^2 -22z +{107 \over 4}-{18 \over z}-{2 \over
1-z}\Bigg)\ln z +14 z^2-{73z \over 8}+{63 \over 4}-{109 \over 8z}
\Bigg\},
\end{eqnarray}
where $N_c=3$. Now the soft divergences coming from virtual gluons
and bremsstrahlung contributions have cancelled exactly according
to the Bloch-Nordsieck theorem \cite{bn}. The remaining
divergences are collinear.

Moreover, calculating the processes with two gluons or two light
quarks in the initial states, we obtain the following partonic
cross sections
\begin{eqnarray} \label{eq:split}
\hat\sigma_{gg}^{\rm bare}(z,1/\epsilon_{IR})
&=&\frac{8\pi\alpha_s^2}{3s} \left(\frac{\kappa_{tc}^g}
{\Lambda}\right)^2 m_t^2 \Bigg[- C_\epsilon^\prime {z^2+(1-z)^2
\over 2} {1 \over \epsilon_{IR}} +\Bigg({z^2 \over 4}+{17z \over
8} -{1 \over 4(1+z)} \nonumber
\\[2ex]
&& +{3 \over 8}  \Bigg)\ln z +(2z^2-2z+1)\ln(1-z)-{259 \over
64}z^2 +{505 \over 128}z+{1 \over 16} +{5 \over 128z}\Bigg],
\nonumber
\\[2ex]
\hat\sigma_{cc}^{\rm bare}(z,1/\epsilon_{IR})
&=&\frac{32\pi\alpha_s^2}{9s} \left(\frac{\kappa_{tc}^g}
{\Lambda}\right)^2 m_t^2 \Bigg[-C_\epsilon^\prime {1+(1-z)^2 \over
z} {1 \over \epsilon_{IR}} +2z\ln(1-z) -z\ln z \nonumber
\\[2ex]
&& -4\ln(1-z)+2\ln z+{10 \over 3}+{4 \over z}\ln(1-z)-{2 \over
z}\ln z- {7 \over 3z}\Bigg], \nonumber
\\[2ex]
\hat\sigma_{c\bar{c}}^{\rm bare}(z,1/\epsilon_{IR}) &=&
\frac{16\pi\alpha_s^2}{9s} \left(\frac{\kappa_{tc}^g}
{\Lambda}\right)^2 m_t^2 \Bigg[-C_\epsilon^\prime{1+(1-z)^2 \over
z} {1 \over \epsilon_{IR}} +2z\ln(1-z) -z\ln z \nonumber
\\[2ex]
&& -{2 \over 3}z-4\ln(1-z)+2\ln z-{4 \over 3z}+{7 \over 3}+{4
\over z}\ln(1-z)-{2 \over z}\ln z+ {2 \over 3}z^2\Bigg], \nonumber
\\[2ex]
\hat\sigma_{cq(\bar q)}^{\rm bare}(z,1/\epsilon_{IR})
&=&\frac{16\pi\alpha_s^2}{9s} \left(\frac{\kappa_{tc}^g}
{\Lambda}\right)^2 m_t^2 \Bigg[-C_\epsilon^\prime {1+(1-z)^2 \over
z} {1 \over \epsilon_{IR}} +2z\ln(1-z) -z\ln z \nonumber
\\[2ex]
&& -4\ln(1-z) +2\ln z+3+{4 \over z}\ln(1-z)-{2 \over z}\ln z- {2
\over z}\Bigg], \nonumber
\\[2ex]
\hat\sigma_{q\bar{q}}(z) &=& \frac{16\pi\alpha_s^2}{9s}
\left(\frac{\kappa_{tc}^g} {\Lambda}\right)^2m_t^2 \left({2 \over
3}z^2-z+{1 \over 3z}\right).
\end{eqnarray}
Note that the divergences appearing in above expressions are
collinear, and the patonic cross section of the $q\bar{q}(q\neq
c)$ initial state is free of singularities.

The bare partonic cross sections in Eqs.~(\ref{eq:gc}) and
(\ref{eq:split}), which contain the collinear singularities
generated by the radiation of gluons and massless quarks, have a
universal structure, and can be factorized into the following form
to all orders of perturbation theory:
\begin{eqnarray}
\hat \sigma_{ab}^{\rm bare}(z,1/\epsilon_{IR}) =\sum_{c,d}
\Gamma_{ca}(z,\mu_F,1/\epsilon_{IR}) \otimes
\Gamma_{db}(z,\mu_F,1/\epsilon_{IR}) \otimes
\hat{\sigma}_{cd}(z,\mu_F)\,, \label{eq30}
\end{eqnarray}
where $\mu_F$ is the factorization scale and $\otimes$ is the
convolution symbol defined as
\begin{eqnarray}
f(z)\otimes g(z)= \int_z^1~{dy \over y} f(y)\,g\left({z \over
y}\right).
\end{eqnarray}
The universal splitting functions
$\Gamma_{cd}(z,\mu_F,1/\epsilon_{IR})$ represent the probability
of finding a particle $c$ with fraction $z$ of the longitudinal
momentum inside the parent particle $d$ at the scale $\mu_F$. They
contain the collinear divergences, and can be absorbed into the
redefinition of the PDF according to mass factorization
\cite{factor1}. Adopting the ${\overline {\rm MS}}$
mass--factorization scheme, we have to $\mathcal{O}(\alpha_s)$
\begin{eqnarray}
\Gamma_{cd}(z,\mu_F,1/\epsilon_{IR}) &=& \delta_{c d} \delta(1-z)-
{1 \over \epsilon_{IR}} {\alpha_s \over 2 \pi}
\frac{\Gamma(1-\epsilon)} {\Gamma(1-2\epsilon)} \left({4\pi\mu_r^2
\over \mu_F^2}\right)^ {\epsilon} P_{cd}^{(0)}(z),
\end{eqnarray}
where $P_{cd}^{(0)}(z)$ are the leading order Altarelli-Parisi
splitting functions \cite{alpha}:
\begin{eqnarray}
P_{qq}^{(0)}(z) &=& \frac{4}{3} \left[\frac{1+z^2}{(1-z)_+}
+\frac{3}{2} \delta(1-z) \right], \nonumber
\\[2ex]
P_{qg}^{(0)}(z) &=& \frac{1}{2}\left[ (1-z)^2 +z^2\right],
\nonumber
\\[2ex]
P_{gq}^{(0)}(z) &=& \frac{4}{3} \frac{(1-z)^2 +1}{z}, \nonumber
\\[2ex]
P_{gg}^{(0)}(z) &=& 2N_c\left[\frac{z}{(1-z)_+} +\frac{1-z}{z}
+z(1-z) \right] +\left( \frac{11}{2} -\frac{n_f}{3}
\right)\delta(1-z).
\end{eqnarray}

After absorbing the splitting functions
$\Gamma_{cd}(z,\mu_F,1/\epsilon_{IR})$ into the redefinition of
the PDFs through the mass factorization in this way, we have the
hard--scattering cross sections $\hat{\sigma}_{ab}(z,\mu_F)$,
which are free of collinear divergences, and depend on the scale
$\mu_F$:
\begin{eqnarray}
\hat\sigma_{gc}(z,\mu_F) &=& \frac{8\pi^2 \alpha_s}{3s}
\left(\frac{\kappa_{tc}^g} {\Lambda}\right)^2m_t^2\delta (1-z)
\nonumber
\\[2ex] &&
+ \frac{4\pi\alpha_s^2}{9s} \left(\frac{\kappa_{tc}^g}
{\Lambda}\right)^2m_t^2 \Bigg \{ 3\ln\frac{m_t^2}{\mu_F^2} \Bigg[
2N_c\left({z \over (1-z)_+}+{1-z \over z} +z(1-z)\right) \nonumber
\\[2ex] &&
+ \left({11 \over 2} - {n_f \over 3}\right) \delta(1-z) +
\frac{4}{3}\left({1+z^2 \over (1-z)_+}+{3 \over
2}\delta(1-z)\right) \Bigg] + \Bigg[\frac{4\pi^2}{3} -15 \nonumber
\\[2ex] &&
-\left (n_f -\frac{29}{2} \right) \ln\frac{\mu_r^2}{m_t^2}
\Bigg]\delta(1-z) +{1 \over 8(1-z)_+}\Bigg( 27z^2 -57z -2(77z +
27)\ln z \nonumber
\\[2ex] &&
-23 -{11 \over z} \Bigg) +\frac{77z+27}{2} \left({\ln(1-z) \over
1-z}\right)_+ -\left(36z^2 -28z+{83 \over 2}-{36 \over
z}\right)\ln(1-z) \nonumber
\\[2ex] &&
+\Bigg(28z^2 -22z +{107 \over 4}-{18 \over z}-{2 \over
1-z}\Bigg)\ln z +14 z^2-{73z \over 8}+{63 \over 4}-{109 \over 8z}
\Bigg\}, \nonumber
\\[2ex]
\hat\sigma_{gg}(z,\mu_F) &=&\frac{8\pi\alpha_s^2}{3s}
\left(\frac{\kappa_{tc}^g} {\Lambda}\right)^2 m_t^2
\Bigg[\ln\frac{m_t^2}{\mu_F^2} {z^2+(1-z)^2 \over 2} +\Bigg({z^2
\over 4}+{17z \over 8} -{1 \over 4(1+z)} \nonumber
\\[2ex]
&& +{3 \over 8}  \Bigg)\ln z +(2z^2-2z+1)\ln(1-z)-{259 \over
64}z^2 +{505 \over 128}z+{1 \over 16} +{5 \over 128z}\Bigg],
\nonumber
\\[2ex]
\hat\sigma_{cc}(z,\mu_F) &=&\frac{32\pi\alpha_s^2}{9s}
\left(\frac{\kappa_{tc}^g} {\Lambda}\right)^2 m_t^2
\Bigg[\ln\frac{m_t^2}{\mu_F^2} {1+(1-z)^2 \over z} +2z\ln(1-z)
-z\ln z \nonumber
\\[2ex]
&& -4\ln(1-z)+2\ln z+{10 \over 3}+{4 \over z}\ln(1-z)-{2 \over
z}\ln z- {7 \over 3z}\Bigg], \nonumber
\\[2ex]
\hat\sigma_{c\bar{c}}(z,\mu_F) &=& \frac{16\pi\alpha_s^2}{9s}
\left(\frac{\kappa_{tc}^g} {\Lambda}\right)^2 m_t^2
\Bigg[\ln\frac{m_t^2}{\mu_F^2} {1+(1-z)^2 \over z} +2z\ln(1-z)
-z\ln z \nonumber
\\[2ex]
&& -{2 \over 3}z-4\ln(1-z)+2\ln z-{4 \over 3z}+{7 \over 3}+{4
\over z}\ln(1-z)-{2 \over z}\ln z+ {2 \over 3}z^2\Bigg], \nonumber
\\[2ex]
\hat\sigma_{cq(\bar q)}(z,\mu_F) &=&\frac{16\pi\alpha_s^2}{9s}
\left(\frac{\kappa_{tc}^g} {\Lambda}\right)^2 m_t^2
\Bigg[\ln\frac{m_t^2}{\mu_F^2} {1+(1-z)^2 \over z} +2z\ln(1-z)
-z\ln z \nonumber
\\[2ex]
&& -4\ln(1-z) +2\ln z+3+{4 \over z}\ln(1-z)-{2 \over z}\ln z- {2
\over z}\Bigg].
\end{eqnarray}

Finally, we combine these finite $\hat{\sigma}_{ab}(z,\mu_F)$ with
the appropriate partonic distribution functions to arrive at the
NLO cross sections for the direct top productions:
\begin{eqnarray}
\sigma^{NLO} &=& \int d x_1 dx_2 \Bigg\{ \Big[G_{g/P_1}(x_1,\mu_F)
G_{c/P_2}(x_2,\mu_F) +(x_1 \leftrightarrow x_2) \Big]
\hat{\sigma}_{gc}(z,\mu_F) \nonumber
\\[2ex] &&
+G_{g/P_1}(x_1,\mu_F) G_{g/P_2}(x_2,\mu_F) \hat{\sigma}_{gg}(z,
\mu_F) +G_{c/P_1}(x_1,\mu_F) G_{c/P_2}(x_2,\mu_F)
\hat{\sigma}_{cc}(z,\mu_F) \nonumber
\\[2ex]
&& + \Big[G_{c/P_1}(x_1,\mu_F) G_{{\bar c}/P_2}(x_2,\mu_F) + (x_1
\leftrightarrow x_2) \Big] \hat{\sigma}_{c\bar{c}}(z,\mu_F)
\nonumber
\\[2ex]
&& + \sum_{q(\bar q)} \Big[G_{c/P_1}(x_1,\mu_F) G_{q(\bar
q)/P_2}(x_2,\mu_F) + (x_1 \leftrightarrow x_2) \Big]
\hat{\sigma}_{cq(\bar q)}(z,\mu_F) \Bigg\}.
\end{eqnarray}

\section{\label{sec:dis}Numerical calculation and discussion}
In this section, we present the numerical results of the NLO QCD
corrections to the direct top quark production via anomalous FCNC
couplings at the Tevatron Run 2 ($\sqrt{S}=2.0~{\rm TeV}$) and the
LHC ($\sqrt{S}=14~{\rm TeV}$). In our numerical calculations, we
take the top quark mass $m_t=178.0$ GeV \cite{par} and the
two--loop evolution of $\alpha_s(\mu_r)$ \cite{runningalphas} with
$\alpha_s(M_Z)=0.118$. Moreover, CTEQ6L (CTEQ6M) PDFs \cite{pdf}
are used in the calculation of the LO (NLO) cross sections. As for
the anomalous couplings, which appear in the expressions as
quadratic factors, we choose $\frac{\kappa_{tq}^g}{\Lambda}=0.01$
TeV$^{-1}$.

In Table~\ref{tab:llk}, we list the LO and NLO cross sections for
the center value of the factorization and renormalization scale
$\mu_F=\mu_r=m_t$. One can see that for $gc\rightarrow t$
($gu\rightarrow t$), the NLO predictions enhance the LO results
61\% (59\%) at the Tevatron Run 2, and the 40\% (52\%) at the LHC,
respectively.

\begin{figure}
%\centerline{\epsfig{file=Graph1.eps,width=500pt}}
\includegraphics[width=0.8\textwidth]{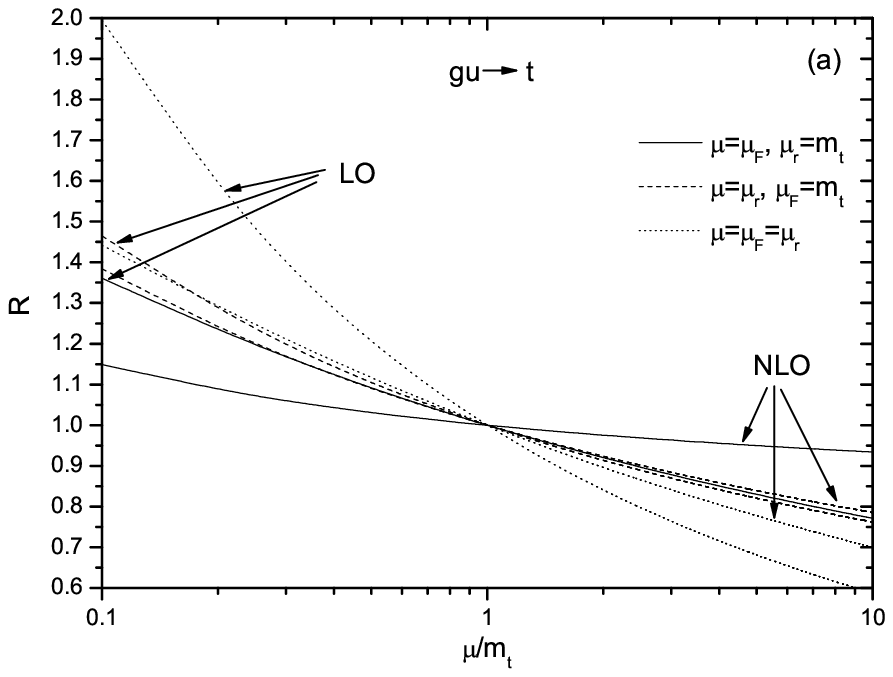}
\includegraphics[width=0.8\textwidth]{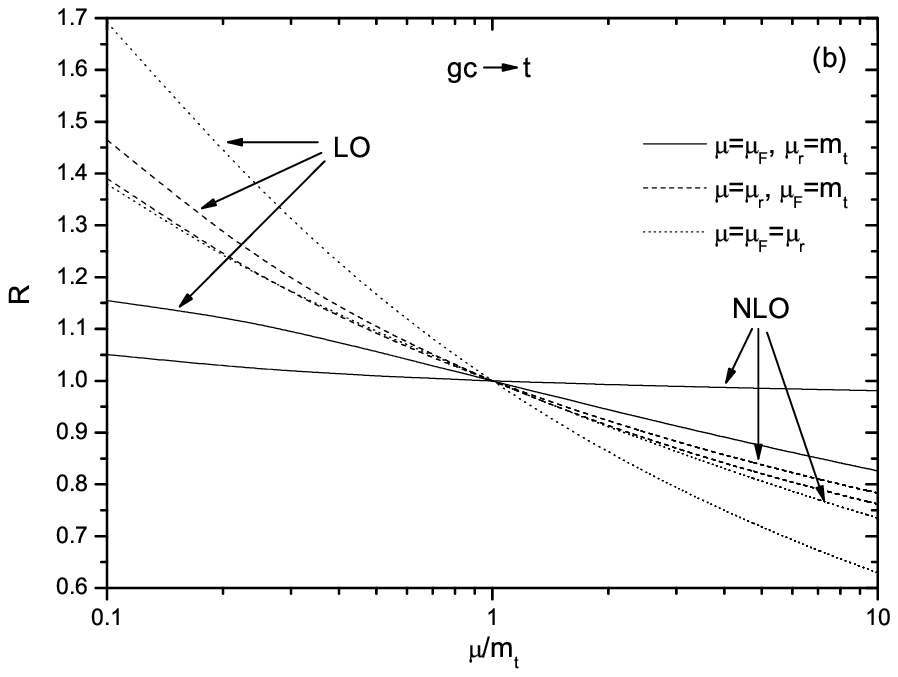}
\caption{The ratio R as functions of factorization
(renormalization) scales for sub-processes $gc\rightarrow t$ and
$gu\rightarrow t$ at the Tevatron Run 2: (a) up quark initial
state and (b) charm quark initial state. Here
$\frac{\kappa_{tc(u)}^g}{\Lambda}=0.01{\rm TeV}^{-1}$.
\label{fig:tevf}}
\end{figure}

We define $R$ as the ratio of the LO (NLO) cross sections to their
values at $\mu_F=\mu_r=m_t$. Fig.~\ref{fig:tevf} shows the ratio R
as functions of factorization (renormalization) scales for
sub-processes $gc\rightarrow t$ and $gu\rightarrow t$ at the
Tevatron Run 2. For each subprocess, we show the scale dependence
for three cases: (1) $\mu_F$ and $\mu_r$ varies simultaneously,
(2) $\mu_F=m_t$ and only $\mu_r$ varies, (3) $\mu_r=m_t$ and only
$\mu_F$ varies. We can find that, for the three cases, the NLO QCD
corrections remarkably reduce the dependence of theoretical
predictions on the factorization (renormalization) scales. For
example, in the region $m_t/2 \le \mu_F=\mu_r \le 2m_t$, the cross
sections vary by $\pm 18 \%$ at LO but by $\pm 11 \%$ at NLO for
$gu\rightarrow t$, and vary by $\pm 15\%$ at LO but by $\pm 9\%$
at NLO for $gc\rightarrow t$.

\begin{figure}
%\centerline{\epsfig{file=Graph2.eps,width=500pt}}
\includegraphics[width=0.8\textwidth]{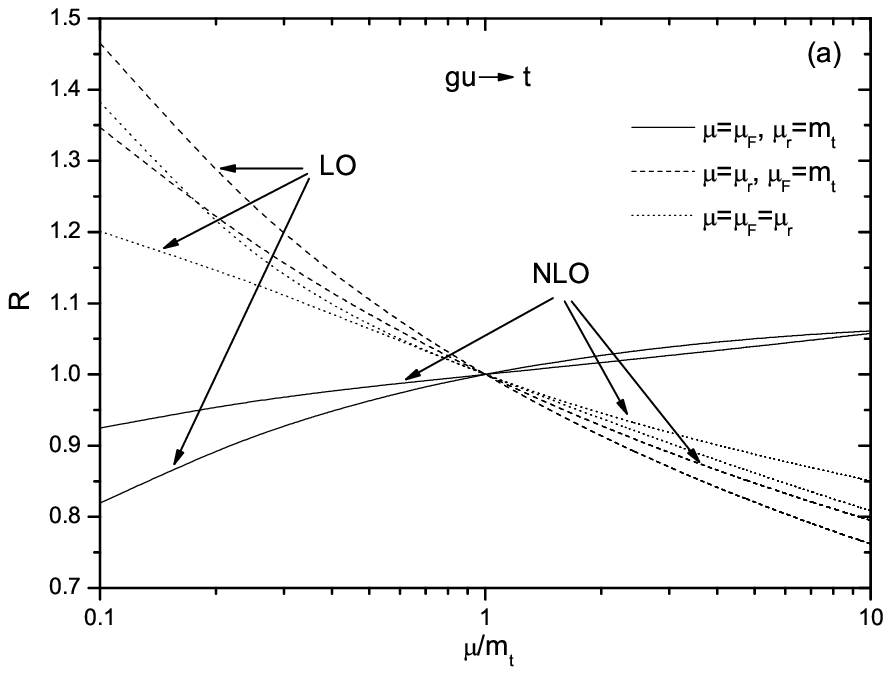}
\includegraphics[width=0.8\textwidth]{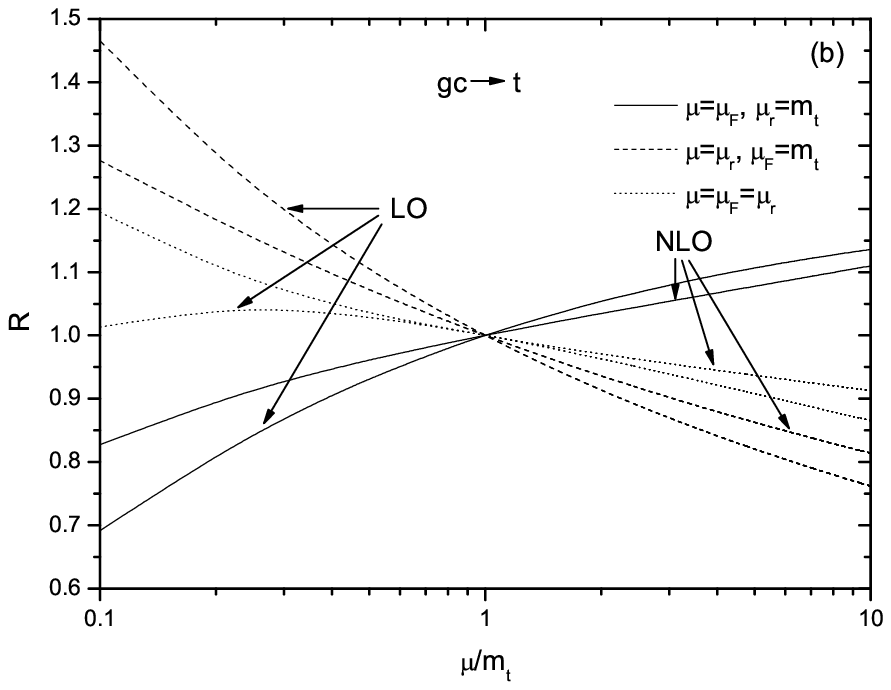}
\caption{The ratio R as functions of factorization
(renormalization) scales for sub-processes $gc\rightarrow t$ and
$gu\rightarrow t$ at the LHC: (a) up quark initial state and (b)
charm quark initial state. Here
$\frac{\kappa_{tc(u)}^g}{\Lambda}=0.01{\rm TeV}^{-1}$.
\label{fig:lhcf}}
\end{figure}

Fig.\ref{fig:lhcf} shows the ratio R as functions of factorization
(renormalization) scales at the LHC. One can find that the scale
dependence is reduced after considering the NLO QCD corrections in
case (2) and (3). For case (1), although the NLO results show the
weaker scale dependence when $\mu_F=\mu_r$ is large ($\ge m_t$),
the NLO results do not reduce the scale dependence of theoretical
predictions for $\mu_F=\mu_r\le m_t$. The main origins to the
scale dependence come from the terms proportional to $\delta(1-z)$
in $\hat\sigma_{gc}(z,\mu_F)$, and higher order effects are
necessary to further reduce such scale dependence in the case of
$\mu_F=\mu_r$ at the LHC. Similar phenomena also can be found in
other processes \cite{ggg}.

\begin{table}
\begin{center}
\begin{tabular}{|c|c|c|c|c|}
\hline
 subprocess & LHC (LO) & LHC (NLO) &
 Tevatron Run 2 (LO) & Tevatron Run 2 (NLO)  \\ \hline \hline
$gu\rightarrow t$ & 11069.8 & 16817.8 & 259.0& 412.6 \\
$gc\rightarrow t$ & 1817.1   & 2536.6  & 17.6 & 28.3 \\
\hline
\end{tabular}
\caption{The LO and NLO cross sections of direct top quark
production via anomalous FCNC couplings at the LHC and Tevatron
Run 2 (fb). Here $\frac{\kappa_{tq}^g}{\Lambda}=0.01$ TeV$^{-1}$
and $\mu_F=\mu_r=m_t$.} \label{tab:llk}
\end{center}
\end{table}

\section{\label{sec:con}Conclusions}

We have calculated the NLO QCD corrections to the cross sections
for direct top quark productions induced by model-independent FCNC
couplings at hadron colliders. The NLO results increase the
experimental sensitivity to the anomalous couplings. Our results
show that the NLO QCD corrections enhance the LO total cross
sections at the Tevatron Run 2 about $60\%$ for both of
$\kappa_{tc}^g$ and $\kappa_{tu}^g$ couplings, and enhance the LO
total cross sections at the LHC about $40\%$ for $\kappa_{tc}^g$
couplings and $50\%$ for $\kappa_{tu}^g$ couplings, respectively.
Moreover, the NLO QCD corrections vastly reduce the dependence of
the total cross sections on the renormalization or factorization
scale, which leads to increased confidence in predictions based on
these results.

\begin{acknowledgments}
We would like to thank C.-P. Yuan and T. Han for useful
discussions. This work was supported in part by the National
Natural Science Foundation of China, Specialized Research Fund for
the Doctoral Program of Higher Education and China Postdoctoral
Science Foundation.
\end{acknowledgments}

\newpage

\newpage

\end{document}